\documentclass[conference]{IEEEtran}
\IEEEoverridecommandlockouts
\pdfoutput=1
\usepackage{cite}
\usepackage{amsmath,amssymb,amsfonts}
\usepackage{algorithmic}
\usepackage{graphicx}
\usepackage{textcomp}
\usepackage{xcolor}
\usepackage{graphicx,import,svg}
\usepackage{pgf}
\usepackage{pgfplots}
\usepackage{wrapfig}
\usepackage{tikz}
\usepackage[RPvoltages]{circuitikz}
\usepackage{subfig}
\graphicspath{{./img/}}
\usepackage[ruled,vlined]{algorithm2e}
\usepackage{algorithmic}

\def\BibTeX{{\rm B\kern-.05em{\sc i\kern-.025em b}\kern-.08em
		T\kern-.1667em\lower.7ex\hbox{E}\kern-.125emX}}
\begin{document}
	
	\title{{XAI-BayesHAR}: A novel Framework for Human Activity Recognition with Integrated Uncertainty and Shapely Values}


	\author{\IEEEauthorblockN{1\textsuperscript{st} Anand Dubey}
		\IEEEauthorblockA{\textit{Infineon Technologies} \\
			Munich, Germany \\
			anand.dubey@infineon.com}
		\and
		\IEEEauthorblockN{2\textsuperscript{nd} Niall Lyons}
		\IEEEauthorblockA{\textit{Infineon Technologies} \\
			Dublin, Ireland \\
			niall.lyons@infineon.com}
		\and

            \IEEEauthorblockN{3\textsuperscript{rd} Avik Santra}
		\IEEEauthorblockA{\textit{Infineon Technologies} \\
			Irvine, CA \\
			avik.santra@infineon.com}
        \and

		\IEEEauthorblockN{4\textsuperscript{th} Ashutosh Pandey}
		\IEEEauthorblockA{\textit{Infineon Technologies} \\
			Irvine, CA \\
			ashutosh.pandey@infineon.com}

	}

	\maketitle
	
	\begin{abstract}
		Human activity recognition (HAR) using IMU sensors, namely accelerometer and gyroscope, has several applications in smart homes, healthcare and human-machine interface systems. In practice, the IMU-based HAR system is expected to encounter variations in measurement due to sensor degradation, alien environment or sensor noise and will be subjected to unknown activities. In view of practical deployment of the solution, analysis of statistical confidence over the activity class score are important metrics. In this paper, we therefore propose \emph{XAI-BayesHAR}, an integrated Bayesian framework, that improves the overall activity classification accuracy of IMU-based HAR solutions by recursively tracking the feature embedding vector and its associated uncertainty via Kalman filter. Additionally, \emph{XAI-BayesHAR} acts as an out of data distribution (OOD) detector using the predictive uncertainty which help to evaluate and detect alien input data distribution. Furthermore, Shapley value-based performance of the proposed framework is also evaluated to understand the importance of the feature embedding vector and accordingly used for model compression. 
	\end{abstract}
	

	\section{Introduction}
	Human activity recognition can assist to automatically save energy in smart homes, such as heating, ventilation, air conditioning and lighting by understanding user's intentions \cite{garg2000,avik2018occ}. Several sensor modalities for human activity recognition are investigated in literature such as camera based, radar \cite{kim2016,javier2014,zenaldin2016,rodrigo2019,vaishnav2020}, Inertial Measurement Unit (IMU) \cite{minarno2020single,masum2019statistical,anik2016activity}, infrared, thermal imaging sensors, etc. The machine learning models used for HAR are optimized using categorical distribution based loss functions such as softmax \cite{kim2016,ha2016convolutional,stadelmayer2020human}.

    Recently, \cite{wan2020deep} evaluated algorithms using CNN, LSTM, BiLSTM, MLP and SVM based architectures for HAR and demonstrated high performance on activity classification by allowing separability of the human activity in the feature space. Although the performance metrics are high for these methods, they fail to provide discrimination between different activity classes in the feature space. This is a critical requirement for HAR system to work in an open world classification setting under variations in the input examples due to sensor noise, or alien environments. Deep metric learning in HAR problems, enables the ability for both separability and discriminating learning by providing distinct class clusters. Contrastive loss \cite{contrastive2014,weiss2018}, triplet loss \cite{hazra2019} and quadruplet loss \cite{chen2017beyond} employ this principle for attaining robust metric based representation learning. 


	However, real-time continuous human sensing and activity classification has challenges arising due to missed missed signals or transitioning between known and unknown activities. To address this, in \cite{dubey2021a,lyons2021improved}, solutions using deep variational embedding models for classification utilizing a tracker based on a Kalman Filter (KF) is introduced. Here, a metric learning based triplet and quadruplet loss function for the optimization of the encoder-decoder architecture is implemented. This helped the network to learn the projection of time series input data into an embedding vector, where data from similar activities are grouped together while dis-similar activities are far apart. During inference, the embedding vector from the learned model is fed into a Kalman filter to track the embedding vector over time, while handling spurious miss-classifications and smoothening the embedding vectors towards an activity cluster centroid. Additionally, the tracked feature embedding is classified into the desired activity class using a K-NN classifier. 
	\begin{figure*}[!htbp]
		\centering
		\scriptsize	
		\def\svgwidth{2.0\columnwidth}
		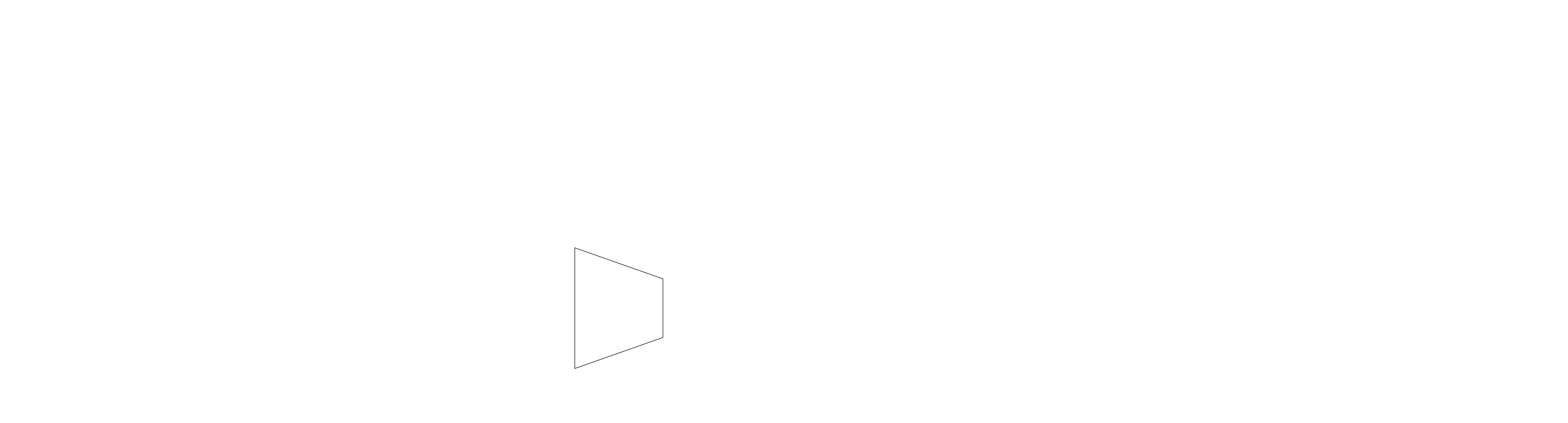
		\vspace{0.1cm}
		\caption{Comparison of (a) State-of-the-art and (b) \emph{XAI-BayesHAR} framework for HAR. In contrast to SOTA, \emph{XAI-BayesHAR} enables the framework to follow complete Bayesian inference over feature extraction and prediction stage. Additionally, \emph{XAI-BayesHAR} uses Shapley value for feature understanding and model compression during training (illustrated in dotted lines).}
		\label{fig:framework}
	\end{figure*}

	While the integrated framework leads to an improved performance in the HAR domain over SOTA \cite{lyons2021improved}, the framework uses K-NN classifier which fails to estimate uncertainty over HAR classification scores. As a result, this leads to further developments in this paper, where we proposed an end to end integrated \emph{XAI-BayesHAR} framework for HAR. For this purpose, an encoder-decoder architecture is trained, similar to \cite{lyons2021improved}, to estimate both mean embedding vector and its associated variance. This helped the integrated tracker to estimate feature vector following Bayesian characteristic. Taking advantage of Bayesian estimate, a 4-layered fully-connected Bayesian neural network (FC-BNN) is proposed to classify the human activity into desired class. FC-BNN leverages the estimate from the tracker by taking both mean and variance as input and thus predicting both classification score with associated uncertainty. Furthermore, Shapley values based concept of feature understanding and selection is applied over learned feature embedding. As a result, this helped to reduce the dimension feature embedding and thus, the size of FC-BNN using iterative algorithm. 

    This paper is organized in 5 section where first section gives introduction to the HAR system and state-of-the-art (SOTA) methods. Later, section 2 gives mathmatical formulation and understanding on \emph{XAI-BayesHAR} followed by training setup in section 3. At the end, both quantitative and qualitative analysis is done in detail. The contributions of our paper are: 
	\begin{itemize} 
		\item The uncertainty associated with activity classification during both training and inference is evaluated to show the reliability and robustness of the framework.
		\item The generalization of the proposed framework is demonstrated by showing its ability to successfully discriminate new unseen highly-correlated targets.
		\item The importance and advantage of interpretation and explainability to reduce the dimension of FC-BNN model.
	\end{itemize}

	\section{Proposed Framework}
	Fig. \ref{fig:framework}(a) illustrates the state-of-the-art (SOTA) pipeline for HAR where an encoder-decoder architecture is optimized in combination with variational approximation together with metric learning at latent space and reconstruction loss at decoder. The learned mean feature embedding from the latent space is then integrated into a K-NN classifier for activity classification. In contrast to this, our framework (termed as \emph{XAI-BayesHAR}), shown in Fig. \ref{fig:framework}(b), follows a encoder-decoder architecture similar to \cite{dubey2021a, lyons2021improved}. This aids the \emph{XAI-BayesHAR} to follow Bayesian characteristic over latent feature embeddings. The \emph{XAI-BayesHAR} feature embedding from the variational encoder is integrated inside the Kalman filter, followed by our proposed FC-BNN network. Due to the inherent nature of Kalman filters and state vector being Bayesian in nature, the tracked and estimated feature embedding also follow Bayesian characteristics. As a result, FC-BNN is optimized within the \emph{XAI-BayesHAR} framework and outputs class scores and its associated uncertainty. The uncertainty quantification not only helps to trust the prediction over a specific target class, but also helps to reject the alien sample treated as out-of-distribution. Finally, a Shapley value based feature understanding is done during the training of the FC-BNN. This helped the network to weight important feature embeddings and thus reduce the size of the FC-BNN to half of original size. The details on \emph{XAI-BayesHAR} integration and formulation are described below. 

	\subsection{ XAI-BayesHAR Integration}
	An encoder is trained to follow variational inference by mapping input data to a distribution over a plausible latent feature embedding. Thus, it returns both mean (confidence $E_{\mu}^{m}$) and variance (uncertainty $E_{\sigma}^{m}$) over the feature embedding. The extracted Bayesian feature embedding is augmented inside the Bayesian KF tracker. The KF assumes the state vector as a Gaussian random variable distribution. Thus, the integration of the latent embedding distribution into the tracker facilitates the processing in obtaining not only the value of the current state of the human activity, but also the uncertainty associated with it. This enables complete Bayesian inference, the estimation of the feature embedding associated with a human activity class. Here, the variance over the embedding vector is used for updating the state uncertainty corresponding to the activity class in the KF. Additionally, due to the nature of the state vector, which carries a probabilistic distribution, a Mahalanobis distance as the association metric is used for the association inside the tracker. This acts as a multivariate Euclidean norm which is a function of both the mean and covariance of the predicted state vector. 
%

	\subsection{FC-BNN Architecture}
	The estimated Bayesian embedding distribution from the tracker is used for HAR. In contrast to the SOTA where K-NN based classifier is applied over mean embedding distribution for activity classification while ignoring variance over it, our proposed FC-BNN takes advantage of Bayesian estimates from the tracker and incorporates input uncertainty in the form of mean and variance. For this purpose, both input and hidden units are designed to follows a Bayesian formulation to predict the categorical distribution, in contrast to the point-estimate neural networks (NNs).

	Although NNs can act as universal approximation function for complex and non-linear functions between inputs and outputs, one of the main limitations of deterministic NNs is that they are fundamentally frequentist in nature. This can be understood from the basic formulation of a cost function during training, i.e., negative log likelihood \footnote{This is equivalent to cross-entropy for categorical and mean square error for Gaussian distribution}. The network is optimized by maximizing likelihood estimates (MLE) over training data $\mathcal{D}=\left\{\left(x_{i}, y_{i}\right)\right\}$ given the network parameters $\boldsymbol{w}.$ Since the network is trying to maximize the probability of data itself, with minimum data, deterministic NNs will often over-fit the data and fail to generalize. As an alternative to this is, instead of calculating the MLE, the maximum a posteriori (MAP) point estimates can be calculated. 
	\begin{equation}
    	\begin{aligned}
    		&\boldsymbol{w}^{\mathrm{MLE}}=\operatorname{argmax}_{\boldsymbol{w}} \log P(\mathcal{D} \mid \boldsymbol{w}) \\
    		&\boldsymbol{w}^{\mathrm{MAP}}=\operatorname{argmax}_{\boldsymbol{w}} \log P(\boldsymbol{w} \mid \mathcal{D})
    	\end{aligned}
            \label{eq:map}
	\end{equation}
	Whereas, both MLE and MAP give point estimates of parameters, they still bring limitation to quantify uncertainty over NN estimates. Thus, this is covered by the posterior predictive distribution, $p(y \mid \mathbf{x}, \mathcal{D})=\int p(y \mid \mathbf{x}, \mathbf{w}) p(\mathbf{w} \mid \mathcal{D}) d \mathbf{w}$ in which the model parameters follow probability distributions and have been marginalized out over parameter $\mathbf{w}$. This helps to reject unseen data which is OOD by quantifying associated uncertainty over its prediction.

	The FC-BNN formulation of the hidden layers and prediction layer helps to propagate both aleatoric and epistemic uncertainty caused due to randomness or error in true estimation from the tracker and lack of model knowledge due to the limited data set \cite{kendall2017uncertainties}. Due to the distribution nature of the model parameters, instead of typical direct backpropagation, these weight distribution parameters are learned through variational inference. This is done by minimizing the Kullback-Leibler (KL) divergence between $q(\mathbf{w} \mid \boldsymbol{\theta})$ and the true posterior $p(\mathbf{w})$ with respect to $\theta$ following below formation.
	\begin{equation}
    	\begin{aligned}
    		\mathcal{F}(\mathcal{D}, \boldsymbol{\theta}) = & \mathrm{KL}(q(\mathbf{w} \mid \boldsymbol{\theta}) \| p(\mathbf{w}))- \\
    		& \mathbb{E}_{q(\mathbf{w} \mid \boldsymbol{\theta})} \log p(\mathcal{D} \mid \mathbf{w}) \\
    		= & \mathbb{E}_{q(\mathbf{w} \mid \boldsymbol{\theta})} \log q(\mathbf{w} \mid \boldsymbol{\theta})-\mathbb{E}_{q(\mathbf{w} \mid \boldsymbol{\theta})} \log p(\mathbf{w})- \\
    		& \mathbb{E}_{q(\mathbf{w} \mid \theta)} \log p(\mathcal{D} \mid \mathbf{w})
    	\end{aligned}
            \label{eq:vi}
	\end{equation}
	As it can be seen from Eq. \ref{eq:vi}, all three formulations are expectation terms with respect to the variational distribution $q(\mathbf{w} \mid \boldsymbol{\theta})$. While the first two terms are data-independent and can be evaluated layer-wise, the last term is data-dependent and is evaluated at the end of the forward-pass. Due to the multi-variant probability distribution nature of the model, it's not possible to compute gradient during backpropagation. Thus, taking advantage of stochastic sampling during forward pass and re-parameterization trick during backward pass, optimization of the network is done. In our case, we initialize model parameters with a Gaussian distribution parameterized by, $\boldsymbol{\theta}=(\boldsymbol{\mu}, \boldsymbol{\sigma})$ where $\boldsymbol{\mu}$ is the mean vector of the distribution and $\sigma$ is the standard deviation vector. For better understanding and generalization of FC-BNN, the network is optimized and analyzed with respect to the state-of-the-art (SOTA) i.e., class feature embedding before integration of KF and our proposed framework i.e., class embedding after temporal smoothening from integrated KF.

	\subsection{Global Interpretability}
	Furthermore, ethical issues surrounding the transparency of data and lack of understanding and trust in machine learning frameworks, creates environments where explainability plays a key role. The ability to interpret and explain why computationally complex models make particular decisions based on given input variables, is inherent to a robust, faithful and trustworthy machine learning platform \cite{agarwal2020interpretable}. Our proposed framework utilizes the SHAP (\textbf{SH}apley \textbf{A}dditive ex\textbf{P}lanations) tool, \cite{lundberg2020shap} based on the game theoretically optimal Shapely Values \cite{rozemberczki2022shapley}. The outcome of this approach illustrates the contribution of each feature on the predicted output. In particular, KernelShap was used throughout to estimate the contributions of each feature value to a prediction. KernelShap improves the sample efficiency of model-agnostic estimations of SHAP values, thus improving the interpretability and explainability of our framework. Additionally, the SHAP values are used in a closed loop with FC-BNN, as illustrated in Fig. \ref{fig:framework}, to downsize both input-dimension and FC-BNN model size. 
	\begin{algorithm}
		\SetAlgoLined
		\SetKwInput{KwData}{Require}
		\KwData{$D^{m}, D^{v}, D^{t}$}
		Baseline Training : \\
		1: Initialize model $\theta^{0}$ \\
		2: $\theta^{m}=\operatorname{train}\left(D^{m}, \theta^{0}, D^{v}\right)$ \\
		3: $Accuracy^{m}=\operatorname{evaluate}\left(\theta^{m}, D^{t}\right)$ \\
		4: Generate SHAP Value \\
		5: maxSHAP = max(SHAP) \\
		6: minSHAP = min(SHAP) \\
		
		7: \While{$\operatorname{test}\left(D^{m}, \theta^{0_{new}}, D^{v}\right) \approx \operatorname{test}\left(D^{m}, \theta^{0}, D^{v}\right)$}{
			8: \textbf{for} ${shap}_{i}$ in $[{shap}_{1}, ,{shap}_{2}, .... {shap}_{n}]$ \textbf{do} \\
			9: $\quad drop(I_{i})$ if $shap^{I_{i}} < SHAP_{th} $\\
			10: \textbf{end for} \\
			11: $\theta^{0}$ = $\theta^{0_{new}}$\\
		}
		\caption{Retraining Procedure using Shapely Values.}
		\label{algo:detection}
	\end{algorithm}
	
	Algorithm \ref{algo:detection} gives an empirical understanding on the FC-BNN model optimization using SHAP value in a closed loop. Here, $D^{m}, D^{v}, D^{t}$ represents the training set, validation set, and the test set used for the optimization of the FC-BNN parameter $\theta^{m}$. After the baseline training of the FC-BNN, SHAP vales are estimated for input feature embeddings. Therefore, based on global maximum-minimum SHAP values, input features ($I_{i}$) having SHAP vales greater than a threshold (SHAP$_{th}$) are selected. In parallel, the hidden units of FC-BNN are also reduced. This is performed in an iterative way until SHAP model accuracy decreases. 
	

	%
	%
	\section{Training Setup}
	\subsection{Dataset Preparation}
	\label{datap}
	The datasets used throughout this paper are considered from \cite{lyons2021improved}. Six Known human activities and one unknown activity are represented using IMU accelerometer and gyroscope sensors, collected by a PSoC 6 Wi-Fi BT CY8CKIT-062 Pioneer Kit and the CY8CKIT-028-TFT shield \cite{singha2019study}. Sensor data was collected at 100Hz, in an indoor setting. Known training data consisted of $21875$ samples, a validation size of $9874$ and a testing size $10197$ samples respectively. Unknown activity samples equated to, $9923$. Time series data were preprocessed using Z-score normalization \cite{patro2015normalization} and a third order Butterworth filter \cite{soi2014design} with a corner frequency of 0.3 Hz to filter noise.

	\subsection{Loss Function}
	\label{qml}
	In the paradigm of metric learning, selection of both triplet and quadruplet pairs are essential. These pairs consist of anchor sample ($x_a$), i.e., any random sample, positive sample ($x_p$), which is from the same class as the anchor, and a negative samples ($x_n$), which is a sample from any different class in comparison to the anchor class. The loss function is computed over feature embedding in the latent space ($z$), which is output from the encoder $q(.)$. Both triplet and quadruplet based optimization is done using online hard and semi-hard pairs following a min-max distance learning between selected pairs, similar to \cite{dubey2021a}.

	Eq.~\ref{eq:loss_triplet} gives mathematical understanding on formulation of triplet loss, where the distance between the anchor and positive samples is minimized forcing $d(q_{\phi}(x_a),q_{\phi}(x_p))$ to $0$ and the distance between the anchor and negative samples is maximized by making $d(q_{\phi}(x_a), q_{\phi}(x_p)) + \alpha_{margin}$ less than $d(q_{\phi}(x_a), q_{\phi}(x_n))$.
	\begin{equation}
		\begin{aligned}
			\mathcal{L}_{\text{triplet}}=&\max(\|q_{\phi}(x_a)-q_{\phi}(x_p)\|^{2}-\|q_{\phi}(x_a)-q_{\phi}(x_n)\|^{2}+\\
			& \alpha_{\text{margin}}, 0), 
			\label{eq:loss_triplet}
		\end{aligned}
	\end{equation}
	Here, $\|.\|$ is the euclidean distance function and $\alpha_{\text{margin}}$ is a hyperparameter, which defines the boundary condition between the similar and dissimilar pairs. In contrast to triplet loss, quadruplet loss includes another negative sample at the cost of another hyperparameter $\alpha_{2}$, summarized by Eq. \ref{eq:loss_qvae}. This helps the network to have a better inter and intra-class distance by adding an extra parameter optimization to separate the negative class from each other. The resulting new loss function is termed as Quadruplet loss ($\mathcal{L}_{\text{quadruplet}}$) and can be summarized by Eq. \ref{eq:loss_qvae}. Here, sample $x_{i}$ and $x_j$ belong to the same class and represent an anchor and positive sample, $x_{k}$ and $x_{l}$ belong to two different classes, which are also not an anchor class.
	\begin{equation}
		\begin{aligned}
			\mathcal{L}_{\text{quadruplet}}	=&\sum_{i, j, k}^{N}\left\|q\left(x_{i}, x_{j}\right)^{2}-q\left(x_{i}, x_{k}\right)^{2}+\alpha_{1}\right\| {+} \\
			& \dot{\sum}_{i, j, k, l}^{N}\left\|q\left(x_{i}, x_{j}\right)^{2}-q\left(x_{l}, x_{k}\right)^{2}+\alpha_{2}\right\| , \\
			& s_{i}=s_{j}, s_{l} \neq s_{k}, s_{i} \neq s_{l}, s_{i} \neq s_{k}
			\label{eq:loss_qvae}
		\end{aligned}
	\end{equation}

	%

	Furthermore, to enable probabilistic inference for the model over learned feature embeddings, the concept of variational inference is adapted which performs an approximate Bayesian inference efficiently by having continuous feature information. The network is optimized by minimizing the upper-bound on the expected negative log-likelihood of the data, together with the desired loss function ($\mathcal{L}_{\text{metric}}$). As a result, in this paper, the overall loss function for both triplet and quadruplet learning can be summarized by Eq. \ref{eq:loss_total}.
	\begin{align}
		\mathcal{L}_{\text{ XAI-BayesHAR}}&= 0.7 * \mathcal{L}_{\text{reconstruction}}+ 0.3 * (\mathcal{L}_{\text{KL}}+\mathcal{L}_{\text{metric}}),
		\label{eq:loss_total}
	\end{align}
	where $L_{\text{KL}}$ is the KL divergence loss at the latent vector to minimize its deviation from Gaussianity with 0 mean and unit variance. $L_{\text{MSE}}$ is the mean-squared error of reconstructing the denoised time series data at each of the decoder, thus termed as reconstruction loss function. And, $\mathcal{L}_{\text{metric}}$ denotes metric loss function, which can be either triplet $\mathcal{L}_{\text{triplet}}$ or quadruplet loss $\mathcal{L}_{\text{quadruplet}}$.

	\section{Results \& Discussion}
	This section investigates on the reliability towards known and unknown target classes using both qualitative and quantitative methods for \emph{XAI-BayesHAR} in comparison to the SOTA. Additionally, the concept of explainability is used to interpret latent embedding vectors and further used for model compression.

	\subsection{Original Feature Embedding}
\begin{figure}[!htbp]
		\centering
		\includegraphics[width=1.0\linewidth]{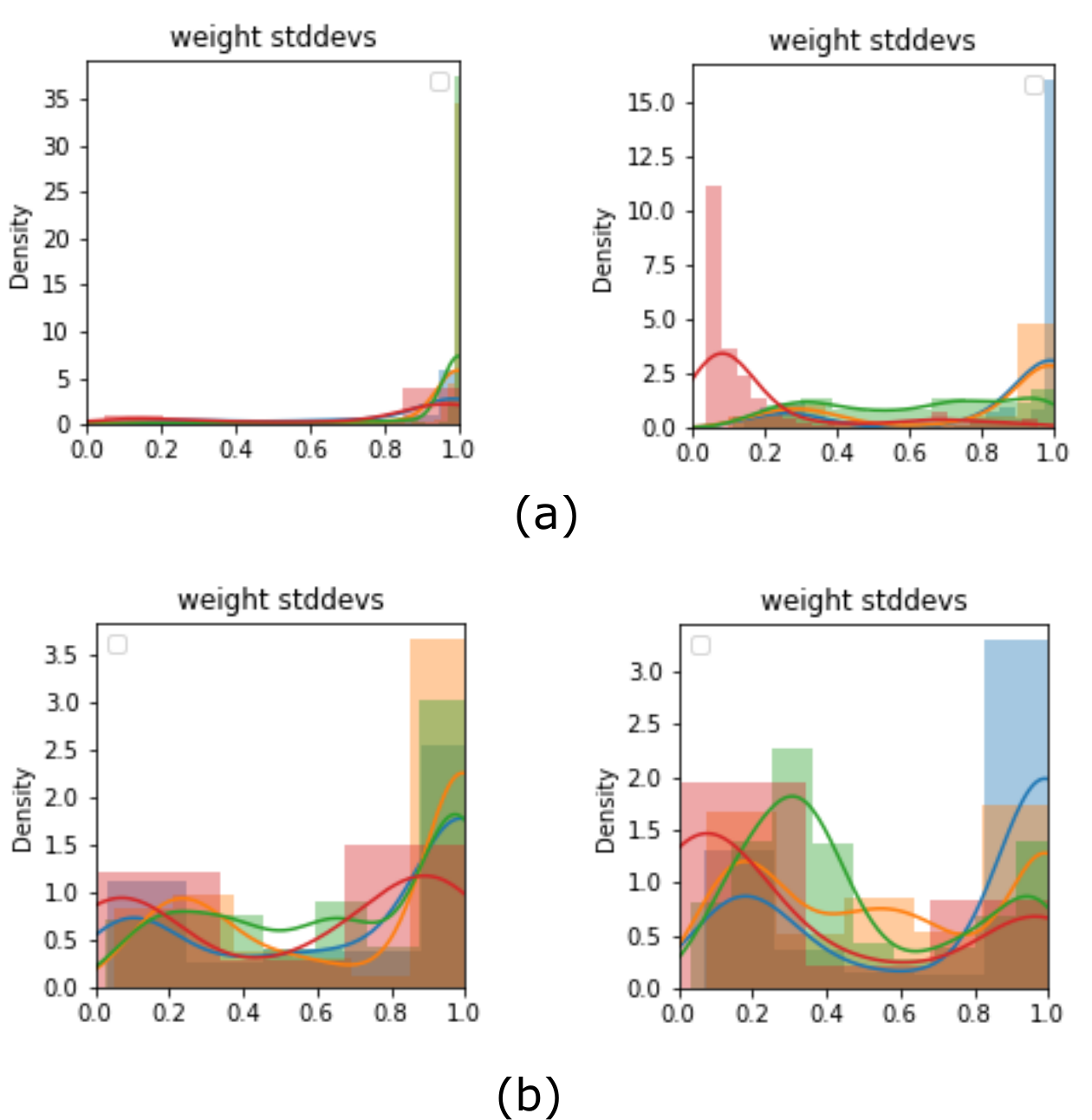}
		\caption{Visualization of model parameter variability distribution for (a) triplet and (b) quadruplet based framework. The left column represents SOTA framework and right column corresponds to \emph{XAI-BayesHAR} where different color represents different depth of the FC-BNN.}
		\label{fig: XAI-BayesHAR_weights}
	\end{figure}

	The variability of pre-trained model parameter distribution is illustrated in Fig. \ref{fig: XAI-BayesHAR_weights}(a) and Fig. \ref{fig: XAI-BayesHAR_weights}(b) for triplet and quadruplet based proposed framework, respectively. 
	\begin{figure}[!ht]
		\centering
		\includegraphics[width=1.0\linewidth]{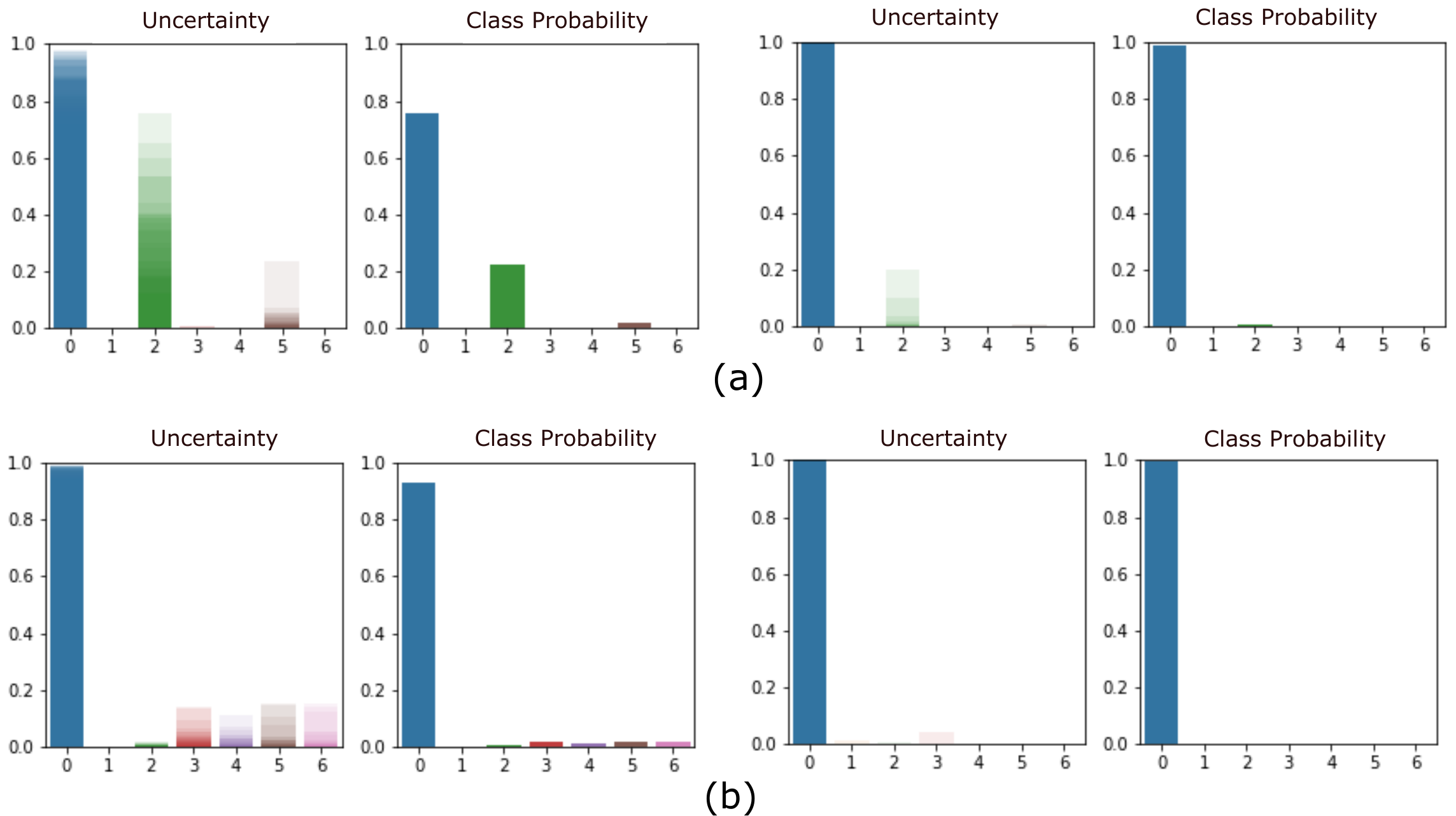}
		\caption{A qualitative analysis of activity classification for class $0$ using FC-BNN for (a) triplet and (b) quadruplet based framework, where left column corresponds to SOTA and right one is for \emph{XAI-BayesHAR}. While both SOTA and \emph{XAI-BayesHAR} classifies activity into true target class, the uncertainty over their prediction for SOTA is much higher in comparison to \emph{XAI-BayesHAR}.}
		\label{fig: XAI-BayesHAR_inlier}
	\end{figure}
	The classification accuracy of SOTA using triplet and quadruplet based optimization is $0.62$ and $0.93$, respectively. The proposed \emph{XAI-BayesHAR} improves classification accuracy to $0.85$ for triplet based framework and $0.97$ for quadruplet based framework. To further understand performance of the different methodologies, variability of learned parameter distribution from FC-BNN is visualized and analyzed in Fig. \ref{fig: XAI-BayesHAR_weights}. The variance over model parameter reflects optimization of the function with varied independent multimode distribution. Thus, a uniform variance characteristic could lead to higher confusion between closely related known or unknown class distribution. Interestingly, the variability of parameter distribution for triplet based SOTA (T-SOTA) in the left column of Fig. \ref{fig: XAI-BayesHAR_weights}(a) is almost flat in contrast to the triplet based proposed framework (T-XAI-BayesHAR), illustrated in the right column in Fig. \ref{fig: XAI-BayesHAR_weights}(a). This gives understanding on poor generalization of the T-SOTA framework in contrast to the T-XAI-BayesHAR for similar or alien human activity target class. In contract to this, SOTA optimized using quadruplet loss (Q-SOTA) shows better variability in parameter distribution in comparison to the T-SOTA, thus results in better performance and generalization. Furthermore, quadruplet based proposed framework (Q-XAI-BayesHAR) enhance the performance of Q-SOTA. This hypothesis is analyzed in the next section using both qualitative and quantitative approach for known activity class data distribution and unknown activity class data distribution.
	
	\begin{figure}[!ht]
		\centering
		\includegraphics[width=1.0\linewidth]{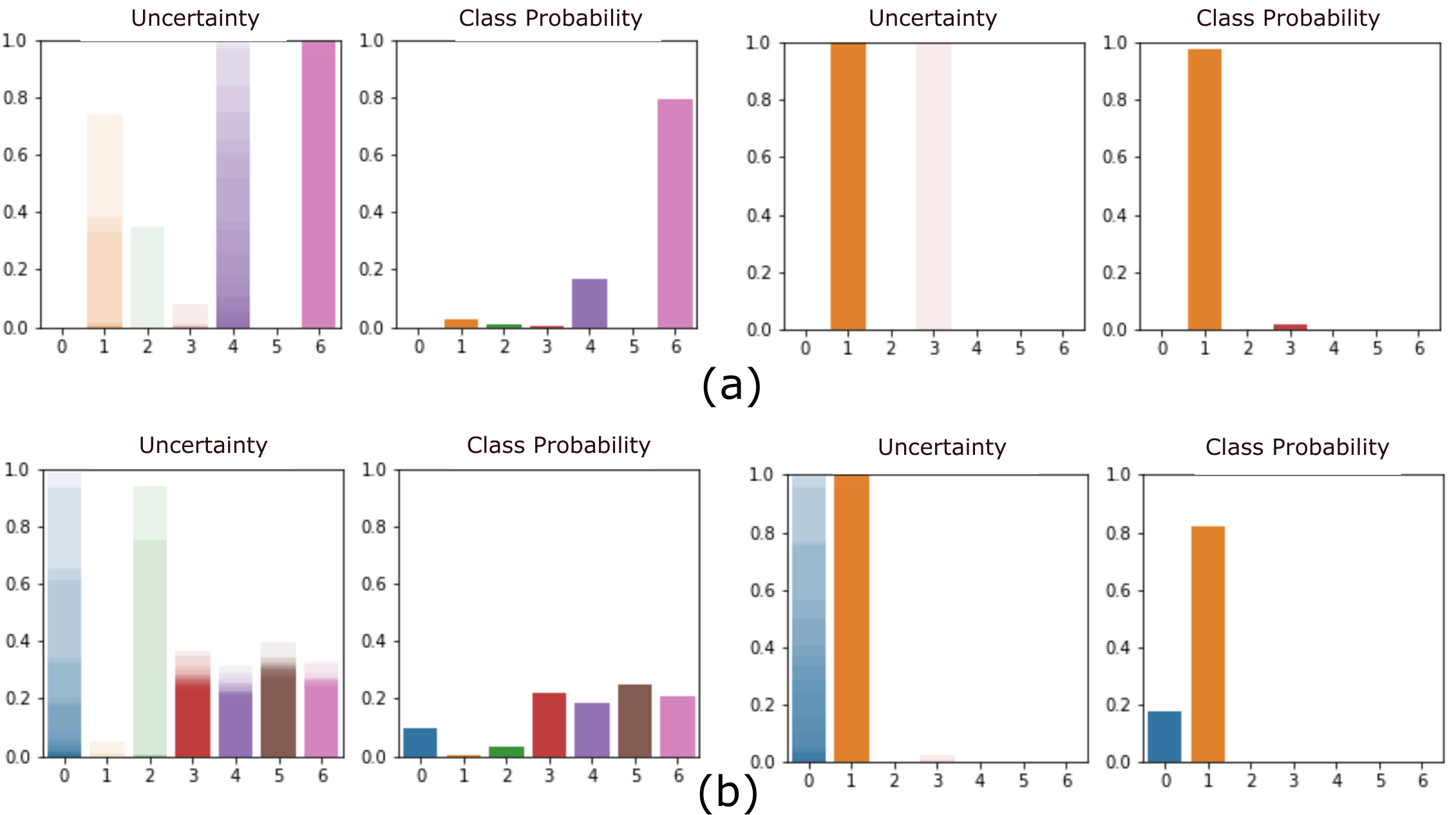}
		\caption{Illustration of qualitative analysis of classification accuracy using pre-trained FC-BNN over unseen activity data-distribution for (a) triplet and (b) quadruplet based framework, where left column corresponds to SOTA and right one is for \emph{XAI-BayesHAR}.}
		\label{fig: XAI-BayesHAR_outlier}
	\end{figure}

	\subsubsection{Known-Data Distribution}
	Fig. \ref{fig: XAI-BayesHAR_inlier} gives a visual understanding of qualitative analysis of activity classification using FC-BNN applied over feature embedding estimated from SOTA (left column) and \emph{XAI-BayesHAR} (right column). While Fig. \ref{fig: XAI-BayesHAR_inlier}(a) corresponds to triplet based optimized encoder, Fig. \ref{fig: XAI-BayesHAR_inlier}(b) represents analysis for quadruplet based optimized encoder. Additionally, in every subplot, the first sub-plot from the left represents class prediction variability (uncertainty) by posterior sample followed by classification score. The x-axis represents human activity (“Idle”: 0, “Jump”: 1, “Sit”: 2, “Squat”: 3, “Stairs”: 4, “Stand”: 5, “Walk”: 6) and the y-axis represents confidence scores in the case of predictive probability and uncertainty in posterior samples.

	It is evident that both SOTA and \emph{XAI-BayesHAR} predicts the true activity class for both the triplet and quadruplet based framework, the T-SOTA shows much higher uncertainty for class $2$ in comparison to the Q-SOTA. This is due to the reason that variability on learned weight parameters for T-SOTA are uniform, as shown in Fig. \ref{fig: XAI-BayesHAR_weights}(a)-left, in contrast to Q-SOTA from Fig. \ref{fig: XAI-BayesHAR_weights}(b)-left. Further, the uncertainty over the prediction estimate is reduced to zero with our proposed integrated framework for T-XAI-BayesHAR and Q-XAI-BayesHAR. Thus, both Fig. \ref{fig: XAI-BayesHAR_inlier} and \ref{fig: XAI-BayesHAR_weights} demonstrates that our framework helps to learn variability among target class distribution while preserving the target class score. In addition to it, this also shows that FC-BNN successfully learns the confusion metric by quantifying uncertainty together with class prediction scores.

	\subsubsection{Out of Distribution}
	Furthermore, \emph{XAI-BayesHAR} is evaluated for a new activity class (kicking), labelled as $7$, which was not seen by both pre-trained encoder and FC-BNN during training. Prior to evaluation of classification scores and associated uncertainty, Pearson coefficient for class $7$ is being calculated against feature embedding estimated from SOTA and \emph{XAI-BayesHAR} for all other human activity, as mentioned in Table \ref{tab2}. 
	\begin{table}[htbp]
		\caption{Pearson coefficient to measure class similarity between known class distribution and out of class distribution for features from SOTA and \emph{XAI-BayesHAR}.}
		\begin{center}
			\begin{tabular}{p{2cm}|p{0.5cm} p{0.5cm} p{0.5cm} p{0.5cm} p{0.5cm} p{0.5cm} p{0.5cm}}
				\hline
				{Unknown Class}&\multicolumn{7}{|c}{Known Class Distribution} \\
				\cline{2-8} 
				{Distribution (7)} & $0$ & $1$ & $2$ & $3$ & $4$ & $5$ & $6$ \\
				\hline
				
				\textit{Q-XAI-BayesHAR} & $-0.03$ & \textbf{\textit{0.52}}& $-0.06$ & $0.47$ & $0.33$ & $0.44$ & $0.43$ \\
				\textit{Q-SOTA} & $\textit{0.49}$ & $\textit{0.99}$ & $\textit{0.47}$ & $\textit{0.78}$ & $\textit{-0.55}$ & $\textit{0.51}$ & $\textit{0.49}$ \\
				\hline
				\textit{T-XAI-BayesHAR} & $\textit{0.24}$ & \textbf{\textit{0.47}} & $\textit{0.29}$ & $\textit{-0.46}$ & $\textit{0.43}$ & $\textit{-0.18}$ & $\textit{0.14}$ \\
				\textit{T-SOTA} & $0.37$ & $0.98$ & $0.29$ & $-0.72$ & $-0.69$ & $-0.31$ & $-0.40$ \\
				\hline
			\end{tabular}
			\label{tab2}
		\end{center}
	\end{table}
	While kicking class $7$ is treated as OOD, the Pearson coefficient for kicking class shows strong relation with jumping class $1$ for both triple and quadruplet based SOTA and \emph{XAI-BayesHAR}. In addition to it, Table \ref{tab2} also shows the decreased correlation between pre-trained class distribution and an alien class for the proposed framework in comparison to SOTA. This further demonstrates better separability of the learned feature embedding using \emph{XAI-BayesHAR} in contrast to the SOTA.
	

	Similar to Fig. \ref{fig: XAI-BayesHAR_inlier}, Fig. \ref{fig: XAI-BayesHAR_outlier} evaluates the performance of the proposed framework as an OOD detector using uncertainty estimates from FC-BNN. For this purpose, the feature embedding of class $7$ activity is extracted from both SOTA and \emph{XAI-BayesHAR} and passed to pre-trained categorical FC-BNN. Due to the inherent nature of categorical based optimization, the network is bound to predict class label with in bounded class distribution (i.e., in our case between $0$ to $6$). By taking advantage of FC-BNN, estimates can be rejected if uncertainty is high.

	As discussed before, due to low separability between pre-trained human activity target class and new alien class, the predictive uncertainty from FC-BNN shows a uniform uncertainty behavior between class $3, 4, 5, 6$ and making it not rely on predictive estimates in the case of SOTA. Furthermore, the feature embedding from class 7 is highly related to other target class, as seen in Table \ref{tab2}. In contrast to this, despite \emph{XAI-BayesHAR} shows better separability between pre-trained target class's, it successfully rejects alien class's with high uncertainty between class $1$ and class $3$ for T-XAI-BayesHAR and class $1$ and class $0$ for Q-XAI-BayesHAR.

	\subsubsection{Local Interpretability}
	\emph{XAI-BayesHAR} utilizes \textbf{SHAP} to interpret the contribution of each feature to the predicted output.
	\begin{figure}[!htb]
		\centering
		\includegraphics[trim={0mm 0mm 0mm 0mm}, clip,width=1.0\linewidth]{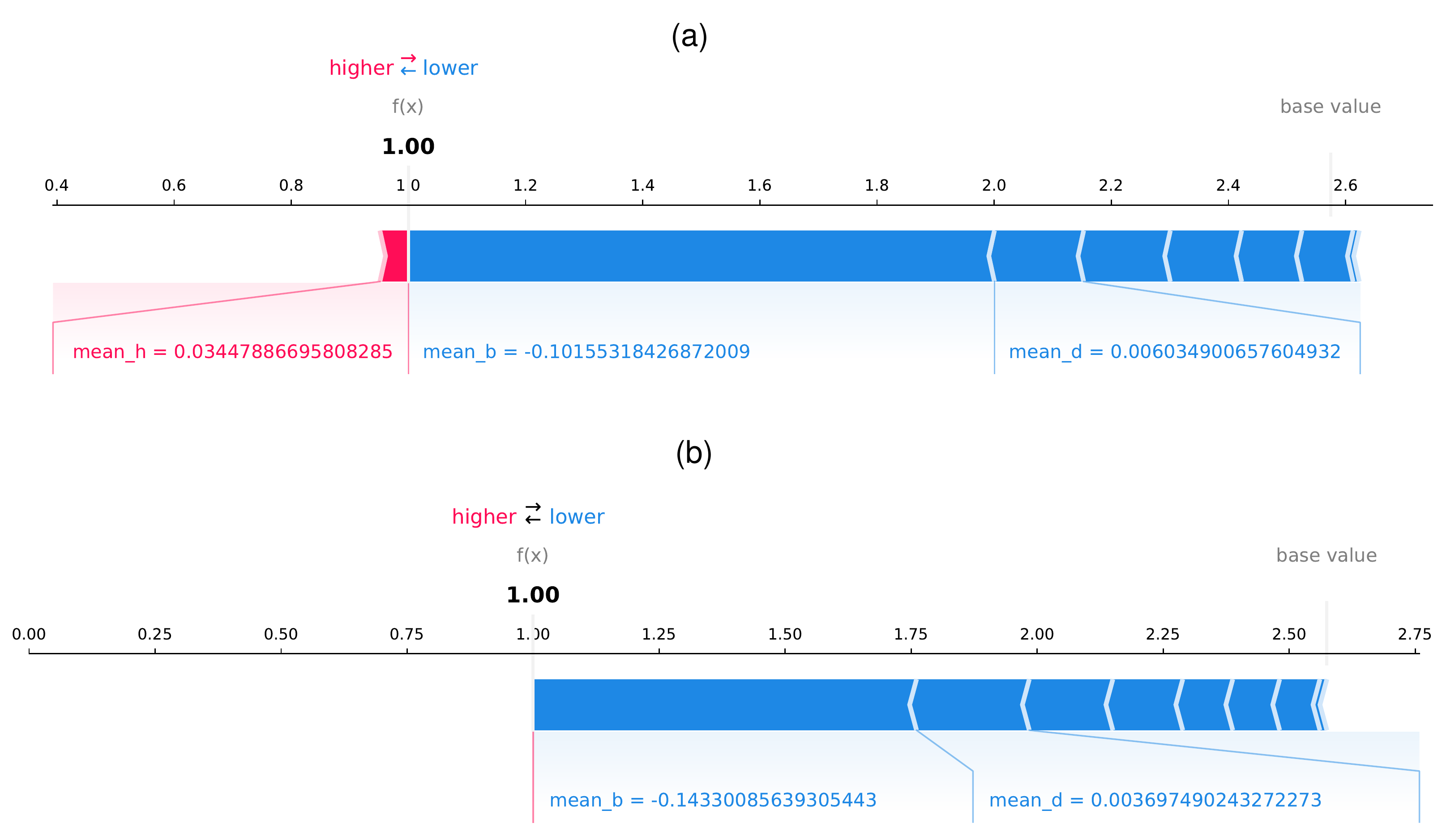}
		\caption{Proposed framework SHAP Force Plot of individual classes (a) Jumping, (b) Kicking}
		\label{fig:shap_force}
	\end{figure}
	As outlined in Table \ref{tab2}, class $7$ (kicking) is treated as out-of-distribution, however the Pearson coefficient shows strong relation with class $1$ (jumping). SHAP force plots are used to see how features contributed to the model’s prediction for a specific observation, namely kicking and jumping. Interestingly, Fig. \ref{fig:shap_force} (a) jumping and (b) kicking share attributes that contribute to the output, \textit{mean\_b} and \textit{mean\_d}. However, it can be seen that the out-of-distribution sample attributes push the prediction value lower, while \textit{mean\_h} in (a) push the prediction value higher. Therefore, although a strong Pearson coefficient, through model interpretability, the \emph{XAI-BayesHAR} successfully demonstrates that the framework can handle classes with a low inter-class difference.

	\subsection{Reduced Feature Embedding}
	In the second stage of our experiment, we further analyze the contribution of each feature embedding and compress the FC-BNN model in a close loop, as mentioned earlier in Algorithm. \ref{algo:detection}. Additionally, uncertainty analysis for both inlier and outlier data distribution is analyzed similar to Fig. \ref{fig: XAI-BayesHAR_inlier} and Fig. \ref{fig: XAI-BayesHAR_outlier}. 
	
\begin{figure}[!htb]
		\centering
		\includegraphics[trim={0mm 0mm 0mm 0mm}, clip,width=1.0\linewidth]{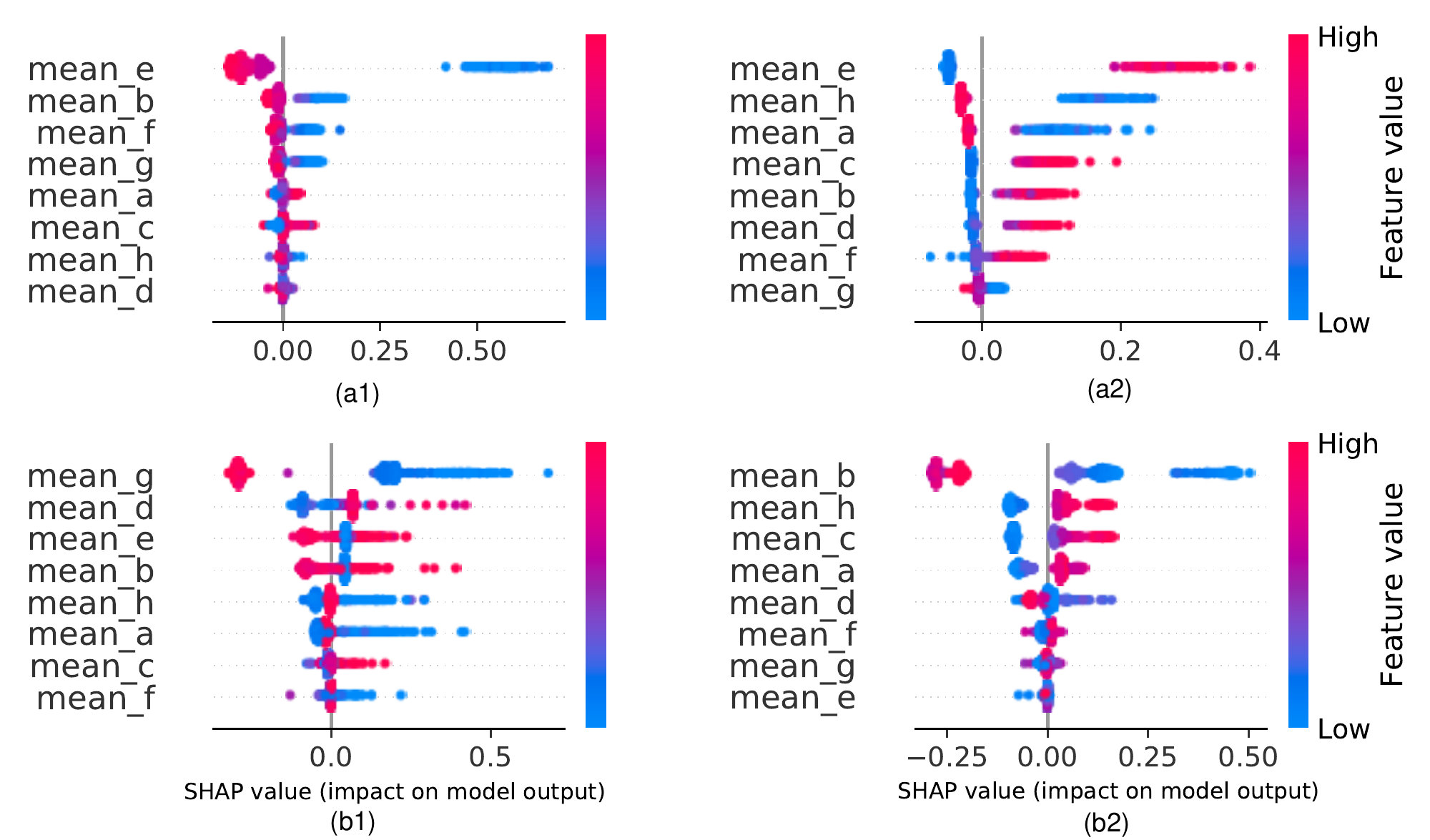}
		\caption{(a) TVAE, and (b) QVAE SHAP values feature summary of SOTA (left) and proposed framework (right)}
		\label{fig:beeswarm}
	\end{figure}

	Fig. \ref{fig:beeswarm} illustrates the contributions made by each feature influence the outcome of the models in ascending order. Both SOTA and \emph{XAI-BayesHAR} frameworks are evaluated for triplet and quadruplet. Each sample in the dataset is represented for each attribute, indicating the importance of that feature on the prediction. The Shap value impact on the horizontal axis of the plot indicates the importance of the prediction. Thus, how each class embedding after temporal smoothening from integrated KF will affect the overall output. The TVAE and QVAE mean output vector, for illustration purposes, is labelled \textit{mean\_a-h}. 

	Interestingly, it can be seen in Fig. \ref{fig:beeswarm} (a1) and (a2) that \textit{mean\_e}, is the most significant feature for the T-SOTA and T-XAI-BayesHAR. However, SHAP value impact in the SOTA is negatively correlated to target classes, while it has a high and positive impact on the target class in the proposed framework. In comparing the QVAE, SOTA and proposed in Fig. \ref{fig:beeswarm} (b1) and (b2) respectively, the SHAP value feature influences changes with the introduction of temporal smoothening. It can be seen that \textit{mean\_b}, \textit{mean\_h}, \textit{mean\_c} has high correlations with the prediction, while \textit{mean\_e} and \textit{mean\_g} have little relevance to the output prediction. Through the addition of exploring models using interpretability and explainable techniques, models are re-trained using knowledge gained from SHAP value plots. Thus, reducing space and time complexities, and creating a noise reduced more interpretable model that is integrated into our closed loop proposed framework.

	\subsubsection{Known-Data Distribution}
\begin{figure}[!htbp]
		\centering
		\includegraphics[width=1.0\linewidth]{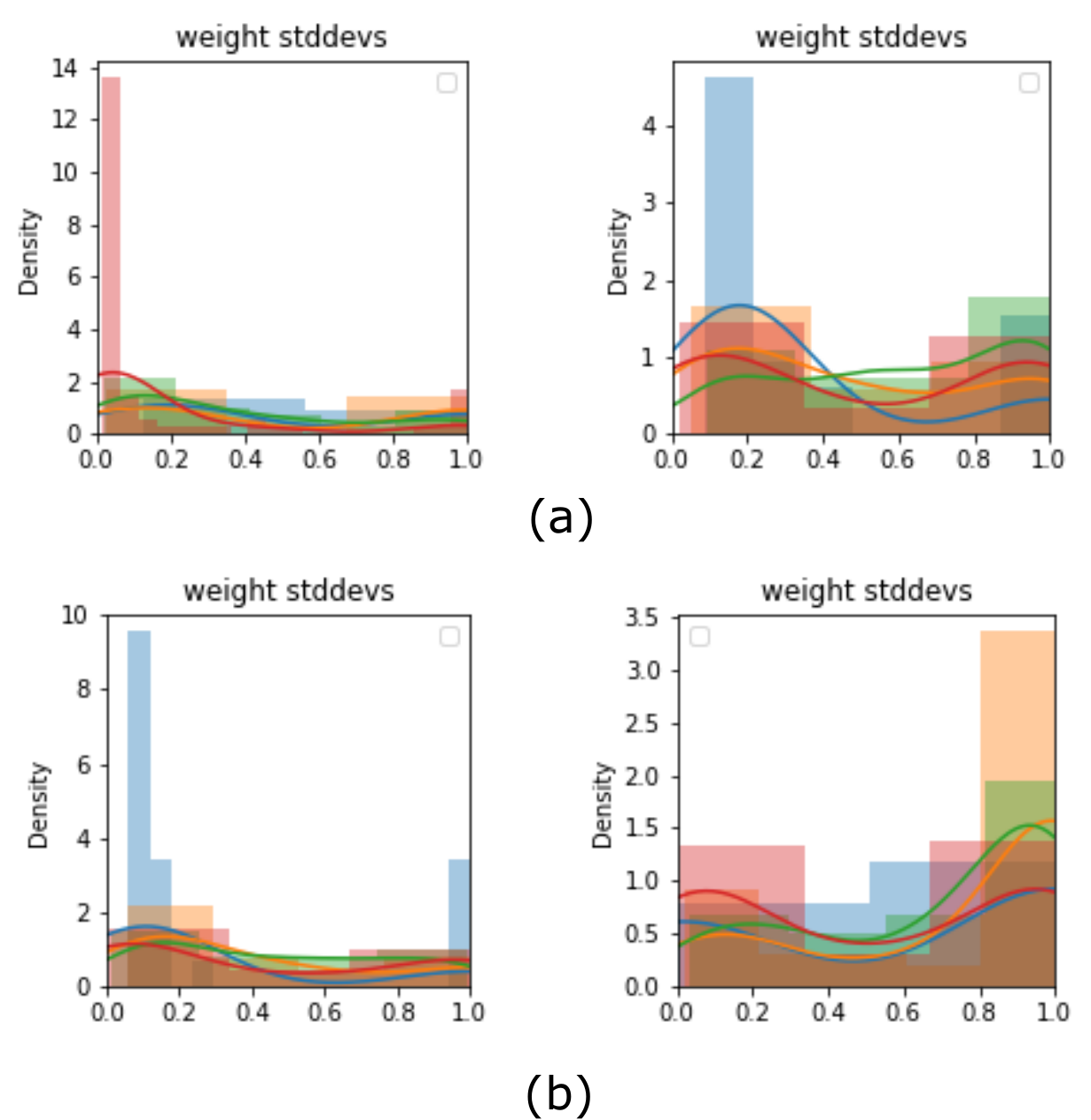}
		\caption{Visualization of model parameter variability distribution for (a) triplet and (b) quadruplet based framework. The left column represents SOTA framework and the right column corresponds to \emph{XAI-BayesHAR} where different color represents different depth of the tiny FC-BNN.}
		\label{fig: XAI-BayesHAR_weights_xai}
	\end{figure}

	Leveraging the SHAP value corresponding to each feature value, the input dimension is reduced to $3$ to that of the original dimension of $16$. This result in the compression of FC-CNN model by a factor of $95\%$  with $402$ trainable parameters in contrast to the original FC-BNN with $7548$ parameters. Following similar procedure as below, the SOTA and \emph{XAI-BayesHAR} give classification accuracy of $0.80$ and $0.95$ for triplet and $0.70$ and $0.82$ for quadruplet based embedding. Interestingly, the triplet based SOTA and \emph{XAI-BayesHAR} shows better classification accuracy in comparison to quadruplet based SOTA and \emph{XAI-BayesHAR}. To better understand this, the tiny FC-BNN classification accuracy is further evaluated using \emph{XAI-BayesHAR} uncertainty. Prior to looking into uncertainty, learned variability inside tiny FC-BNN is evaluated and illustrated in Fig. \ref{fig: XAI-BayesHAR_weights_xai}, similar to Fig. \ref{fig: XAI-BayesHAR_weights}. The Fig. \ref{fig: XAI-BayesHAR_weights_xai} shows uniform learned parameter variability for the case of both T-SOTA and Q-SOTA. This gives an indication for over-fitting of tiny FC-BNN model using reduced triplet embedding and poor separability of different activity class. 
	\begin{figure}[!ht]
		\centering
		\includegraphics[width=1.0\linewidth]{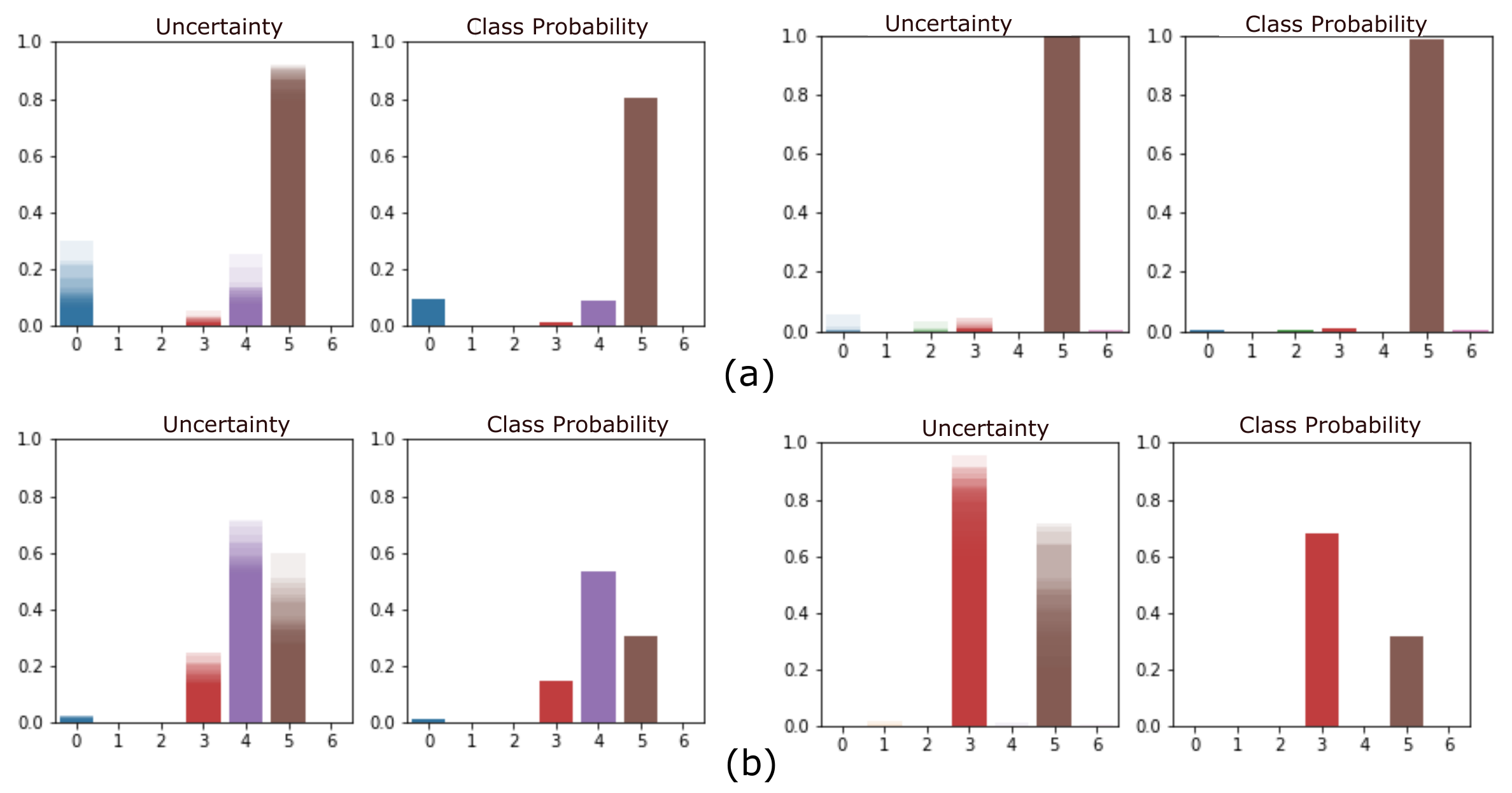}
		\caption{A qualitative analysis of pre-trained FC-BNN over unseen activity data-distribution estimated from (a) SOTA and (b) our \emph{XAI-BayesHAR} framework. The uncertainty from \emph{XAI-BayesHAR} is much higher in comparison to SOTA and help to discard the prediction.}
		\label{fig: XAI-BayesHAR_inlier_xai}
	\end{figure}

	Further, Fig. \ref{fig: XAI-BayesHAR_inlier_xai} shows the behavior of tiny FC-BNN over reduced feature embeddings for known activity class's. Clearly, SOTA framework shows much higher uncertainty between multiple target class. This is due to reason of uniform variability in the learned model parameters inside the tiny FC-BNN. Additionally, T-SOTA shows higher confidence over the prediction class in comparison to the Q-SOTA. The similar trend is followed within \emph{XAI-BayesHAR} whereas, advantage of our \emph{XAI-BayesHAR} can be reflected on increased uncertainty over target class 3 in case of false prediction.

	\subsubsection{Unknown-Data Distribution}
	In addition to uncertainty over reduced embedding for learned activity class embedding, both SOTA and \emph{XAI-BayesHAR} is evaluated for reduced dimension feature embedding for alien class $7$. Table \ref{tab3} shows that the correlation between the alien class and known activity class for reduced feature embedding follows the similar pattern as to the original dimension of feature embedding described in Table \ref{tab2}. 
	\begin{table}[htbp]
		\caption{Pearson coefficient to measure class similarity between known class distribution and out of class distribution for features before integration of KF and after temporal smoothening with KF}
		\begin{center}
			\begin{tabular}{p{2cm}|p{0.5cm} p{0.5cm} p{0.5cm} p{0.5cm} p{0.5cm} p{0.5cm} p{0.5cm}}
				\hline
				{Unknown Class}&\multicolumn{7}{|c}{Known Class Distribution} \\
				\cline{2-8} 
				{Distribution (7)} & $0$ & $1$ & $2$ & $3$ & $4$ & $5$ & $6$\\
				\hline
				\textit{Q-\emph{XAI-BayesHAR}} & $\textit{0.13}$ & \textbf{\textit{0.86}} & $\textit{0.32}$ & $\textit{0.26}$ & $\textit{-0.28}$ & $\textit{-0.58}$ & $\textit{-0.68}$ \\
				\textit{Q-SOTA} & $0.6$ & $0.86$ & $0.58$ & $-0.67$ & $-0.64$ & $-0.65$ & $-0.68$\\
				\hline
				\textit{T-XAI-BayesHAR} & \textbf{\textit{0.86}} & $\textit{-0.01}$ & $\textit{0.25}$ & $\textit{-0.26}$ & $\textit{-0.28}$ & $\textit{0.14}$ & $\textit{0.5}$ \\
				\textit{T-SOTA} & $0.99$ & $0.22$ & $0.67$ & $-0.56$ & $-0.62$ & $-0.17$ & $0.60$ \\
				\hline
			\end{tabular}
			\label{tab3}
		\end{center}
	\end{table}
	\begin{figure}[!htbp]
		\centering
		\includegraphics[width=1.0\linewidth]{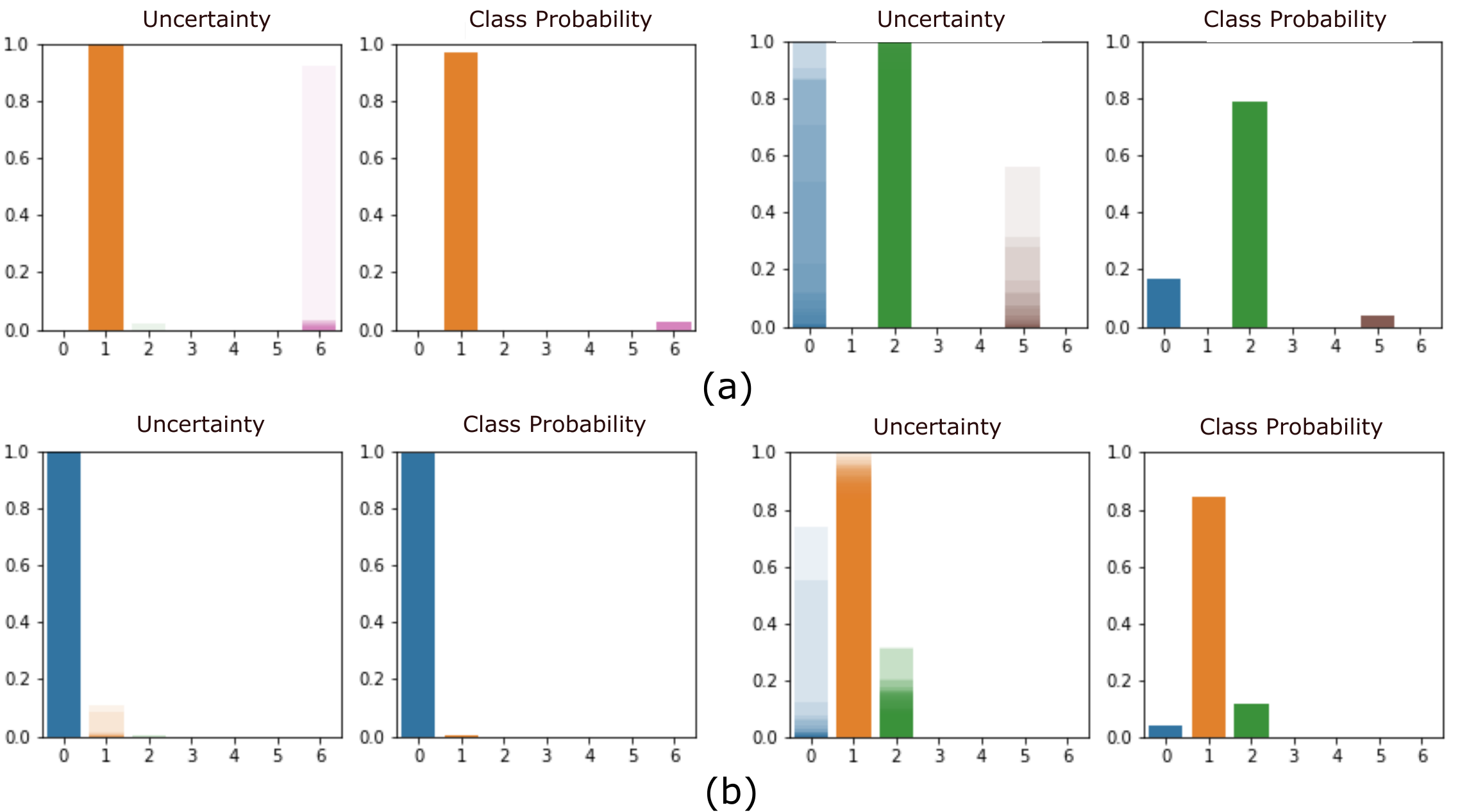}
		\caption{A qualitative analysis of pre-trained FC-BNN over unseen activity data-distribution estimated from (a) SOTA and (b) our \emph{XAI-BayesHAR} framework. The uncertainty from \emph{XAI-BayesHAR} is much higher in comparison to SOTA and helps to discard the prediction.}
		\label{fig: XAI-BayesHAR_outlier_xai}
	\end{figure}

	Fig. \ref{fig: XAI-BayesHAR_outlier_xai} shows the behavior of tiny FC-BNN over reduced feature embeddings for alien data. Clearly, both T-SOTA and Q-SOTA fail to generalize over new unseen data in reduced feature embedding and leads to failing to identify outlier class activity. Whereas, both T-XAI-BayesHAR and Q-XAI-BayesHAR shows high uncertainty over target class $0$, $2$ and $5$ for T-XAI-BayesHAR, and $0$ and $1$ for Q-XAI-BayesHAR. This indicates that despite quadruplet based optimization for both SOTA and \emph{XAI-BayesHAR} over reduced feature embeddings, it shows reduced accuracy in comparison to the triplet based framework. In contrast to the quadruplet based framework, which demonstrates better separability and robustness against outliers while reducing model dimension.

	\section{Conclusion}
	Our proposed framework, \emph{XAI-BayesHAR}, enables a complete Bayesian formulation for HAR using IMU sensors. For this purpose, \emph{XAI-BayesHAR} takes advantage of Bayesian formulation over both feature learning and activity classification. As a result, the framework shows robustness against alien target class (kicking in our case) and rejects it with high uncertainty over classification scores. Additionally, the metric based learning and temporal smoothening used inside \emph{XAI-BayesHAR} further improves the classification in comparison to the SOTA by $\approx 35$\% and $\approx 5$\% for triplet and quadruplet based \emph{XAI-BayesHAR}, respectively. Furthermore, the concept of SHAPLEY value based model explainability is used in a closed loop to compress the FC-BNN by a factor of $95\%$. The tiny FC-BNN successfully retains the separability of both known and unknown activity class applied over \emph{XAI-BayesHAR} in comparison to the SOTA.  

	\bibliographystyle{./bibliography/IEEEtran}
	\bibliography{./bibliography/icmla22}
	
\end{document}